\documentclass[epsfig, 12pt, onecolumn]{article}
\usepackage{epsfig}

\newcommand{\newc}{\newcommand}
\newc{\gsim}{\lower.7ex\hbox{$\;\stackrel{\textstyle>}{\sim}\;$}}
\newc{\lsim}{\lower.7ex\hbox{$\;\stackrel{\textstyle<}{\sim}\;$}}
\newc{\gev}{\,{\rm GeV}}
\newc{\mev}{\,{\rm MeV}}
\newc{\ev}{\,{\rm eV}}
\newc{\kev}{\,{\rm keV}}
\newc{\tev}{\,{\rm TeV}}

\newc{\mz}{M_Z}
\newc{\mpl}{M_*}
\newc{\mw}{m_{\rm weak}}
\newc{\nr}[1]{N^c_R{}_{#1}}

\def\beq{\begin{equation}}
\def\eeq{\end{equation}}
\def\bea{\begin{eqnarray}}
\def\eea{\end{eqnarray}}
\def\bitem{\begin{itemize}}
\def\eitem{\end{itemize}}
\newc{\ie}{{\it i.e.}}          \newc{\etal}{{\it et al.}}
\newc{\eg}{{\it e.g.}}          \newc{\etc}{{\it etc.}}
\newc{\cf}{{\it c.f.}}
\def\vev#1{\left\langle #1 \right\rangle}

\def\inv{^{\raise.15ex\hbox{${\scriptscriptstyle -}$}\kern-.05em 1}}
\def\lbar{{\lower.35ex\hbox{$\mathchar'26$}\mkern-10mu\lambda}} %lambda bar

\let\<=\langle
\let\>=\rangle

\let\+=\uparrow
\let\al=\alpha
\let\be=\beta

\let\ep=\epsilon

\let\th=\theta

\begin{document}
\thispagestyle{empty}
\vspace*{.5cm}
\noindent
\hspace*{\fill}{\large OUTP-04/08}\\
\vspace*{2.0cm}

\begin{center}
{\Large\bf A Simple Model of Neutrino Masses from Supersymmetry Breaking}
\\[2.5cm]
{\large John March-Russell and Stephen West
}\\[.5cm]
{\it Theoretical Physics, Department of Physics\\
University of Oxford, 1 Keble Road, Oxford OX1 3NP, UK}
\\[.2cm]
(February, 2004)
\\[1.1cm]

{\bf Abstract}\end{center}
\noindent
We analyze a class of supersymmetric models first introduced by
Arkani-Hamed et al and Borzumati et al in which the light neutrino masses
result from higher-dimensional supersymmetry-breaking terms in the
MSSM super- and Kahler-potentials.
The mechanism is closely related to the Giudice-Masiero
mechanism for the MSSM $\mu$ parameter, and leads
to TeV-scale right-handed neutrino and sneutrino states,
that are in principle accessible to direct experimental study.
The dominant contribution to the light neutrino (Majorana) mass matrix
is a one-loop term induced by a lepton-number violating $B$-term
for the sneutrino states that is naturally present.  We focus
upon the simplification and analysis of the flavour structure
of this general class of models, finding that simple
and novel origins for the light neutrino mass matrix
are possible.  We find that a subdominant tree-level `see-saw'
contribution may lead to interesting perturbations of the leading
one-loop-induced flavour structure, possibly generating the small
ratio $\Delta m_{\rm solar}^2/\Delta m_{\rm atm}^2$ dynamically.

\newpage

\setcounter{page}{1}

\section{Introduction}

The case for the existence of small neutrino masses and associated
physical neutrino mixing angles has enormously strengthened in
recent years as a consequence of the now numerous experimental
studies of atmospheric and solar neutrinos, and neutrinos from
terrestrial sources. In particular, a recent global analysis
\cite{Maltoni:2004ex} which incorporates data from CHOOZ
\cite{Apollonio:1999ae}, SNO \cite{Ahmed:2003kj}, KamLAND
\cite{Eguchi:2002dm} and Super-Kamiokande \cite{Toshito:2001dk},
leads to values for the mass difference squares, $\Delta m^2$, and
three real mixing angles $\theta_{ij}$ of the neutrinos in the
3$\sigma$ ranges, $1.4\times 10^{-3}\ev^2 \leq|\Delta m^2_{23}|
\leq 3.7\times 10^{-3} \ev^2$ and $0.36 \leq \sin^2 \theta_{23}
\leq 0.67$ from the atmospheric data, and $5.4\times 10^{-5} \ev^2
\leq \Delta m^2_{21} \leq 9.5\times 10^{-5} \ev^2$ and $0.23 \leq
\sin^2\theta_{21} \leq 0.39$ from the solar data, while the
remaining real mixing angle is bounded by $\sin^2\theta_{13}\leq
0.066$. The phases that can appear in the neutrino mixing (MNS)
matrix are currently unconstrained.

The traditional and much studied explanation for
these small masses is the see-saw mechanism\cite{seesaw}.  This
hypothesises the existence of two or more standard-model-singlet
right-handed (rhd) neutrino states $N_i$,
with very large lepton-number violating Majorana masses,
which couple to the weak-SU(2) lepton doublets $L_j$ via a conventional
Yukawa coupling $\lambda_{ij}L_iN_j H$ involving the electroweak Higgs.
The Yukawa couplings
are typically taken to be of size comparable to that of either the charged
lepton Yukawas or the quark Yukawas, depending on the precise
model.  After the Higgs gains its vacuum expectation value, $v$, this
Yukawa coupling leads to the Dirac mass matrix $m_\nu^D=\lambda v$.
Finally, the light neutrino mass matrix obtained by integrating
out the heavy rhd neutrinos is given by
\beq
m_\nu = - (m_{\nu}^D)^T M_R^{-1} m_\nu^D,
\eeq
where $M_R$ is the rhd Majorana mass matrix. If we take the
heaviest Dirac mass to be of order either the $\tau$ lepton or bottom quark
mass, then
agreement with the light neutrino mass inferred from the atmospheric
neutrino anomaly requires roughly $M_R \sim 10^{11}\gev$, while if
the heaviest Dirac mass is taken to be of order the top quark mass, as is commonly
the case in explicit models, then
$M_R\sim 10^{15}\gev$.  The comparative proximity of this second scale
to the inferred supersymmetric grand unified (GUT) scale
$M_{\rm GUT} \simeq 2\times 10^{16}\gev$ is interpreted as evidence in favour
of the see-saw mechanism, as is the fact that rhd
SU(3)$\times$SU(2)$\times$U(1) singlet neutrinos are naturally
included in many GUT theories.

Certainly therefore the see-saw mechanism is an attractive explanation
of why the light neutrino masses are so small.  However, it is not
without its faults.  In particular there is a tension between the
strongly hierarchical nature of the observed Yukawa couplings in the
quark and charged lepton sectors, and the essentially hierarchy-free
masses implied by the $\Delta m^2$'s.  Moreover, both the $\theta_{12}$ and
$\theta_{23}$ mixing angles are large while the angle $\theta_{13}$
is small which is in sharp contrast with the corresponding mixings in
the quark sector which are all small.  These problems can be solved in
specific models, for example the $\Delta m^2$ values can be fitted
by taking the spectrum of rhd neutrino
masses to be hierarchical in such a way as to almost compensate for the
hierarchical neutrino Yukawa couplings.  But this has the price of
introducing a wide range of rhd neutrino masses
$M_R \sim 10^{10} - 10^{15}$ which then require explanation.

In addition, there is the worrying question of the testability of the
see-saw mechanism.  Given the large mass scale of the rhd neutrino
states there is very little prospect of their ever being a {\em direct}
test of the correctness of the see-saw mechanism.  We are forced to fall
back on indirect, and sadly not definitive tests.  For example, the
discovery of neutrino-less
double beta decay would point to a Majorana nature for the light neutrinos
as required by the see-saw mechanism, but as we shall see below, the
see-saw mechanism is by no means the only way that such Majorana masses
can be generated.

In this letter we study a notably attractive alternative,
advocated by Arkani-Hamed et
al~\cite{Arkani-Hamed:2000bq,Arkani-Hamed:2000kj}, and Borzumati
et al~\cite{Borzumati:2000mc,Borzumati:2000ya} (for related work
see \cite{related}),
that links the light neutrino masses to TeV-scale supersymmetry
breaking physics, and which has the significant virtue, compared
to the see-saw mechanism, of being directly testable, at least in
part, at the LHC and other proposed high-energy colliders.  In
overall structure our models are similar to those previously
studied in Refs.\cite{Arkani-Hamed:2000bq, Arkani-Hamed:2000kj,
Borzumati:2000mc,Borzumati:2000ya}, in that the dominant
contribution to the light neutrino masses arises from a one-loop
diagram involving a supersymmetry breaking and lepton-number
violating $B$-term for the right handed sneutrinos, but, by an
alteration of the model, we have been able to significantly
simplify the way in which the flavour structure of the light
neutrino mass matrix arises, and are thus able for the first time
to study in detail some of the consequences of this very
attractive class of models.

\section{Outline of the model}

Supersymmetric extensions of the standard model, such as
the minimal supersymmetric standard model (MSSM), explain the
origin of the weak scale in terms of the scale of the coefficients of the
supersymmetry breaking soft operators that must be included within the
MSSM.  The usual assumption is that these susy-breaking
operators arise from the (super)gravitational mediation
of susy breaking that occurs primordially in some hidden sector at an
intermediate scale $m_I\sim 10^{10}-10^{11}\gev$, giving rise to soft susy
breaking in the MSSM of order $\tev \sim m_I^2/M$.  Here $M$ is the
reduced Planck mass $M = M_{\rm pl}/\sqrt{8\pi} = 2\times 10^{18}\gev$.
Infamously, there is one mass parameter in the MSSM, the $\mu H_u H_d$
superpotential interaction, that naively appears to be independent of
supersymmetry breaking.  If this were indeed true then we would lose
completely our understanding of the origin of the weak scale in susy
theories \cite{Nilles:1983ge}.  The realization of
Giudice and Masiero \cite{Giudice:1988yz} was that the potentially
Planck-scale $\mu H_u H_d$ term in
the superpotential can be forbidden by a global symmetry, while an
effective $\mu$-term is generated from $1/M$-suppressed
terms in the Kahler potential via supersymmetry breaking
effects, thus naturally implying $\mu\sim\tev$.

As emphasized by Arkani-Hamed et al and Borzumati et al, the lesson of the
Giudice-Masiero mechanism for neutrino masses
in the context of susy theories is that SM-singlet operators, such
as the rhd neutrino mass $M_R N N$, or the neutrino Yukawa
coupling $\lambda L N H_u$, might only appear to be renormalizable
superpotential terms, but in fact may arise from $1/M$-{\em suppressed
terms involving the supersymmetry breaking scale} $m_I$.

Specifically, consider the usual MSSM Lagrangian to be supplemented by
a set of superpotential and Kahler terms involving the rhd neutrino
superfields $N_i$ ($i=1,2,3$ is a generation index), and two
Standard-Model-singlet chiral superfields $X$ and $Y$ which arise from the
hidden sector.   In general, the fields which communicate
supersymmetry breaking to the neutrinos can either be flavour singlets
or flavour non-singlets.  Let us generically
call the flavor non-singlet fields, $X$, (which therefore carry
generation indices $i,j$), and the flavour singlet fields, $Y$, and
for simplicity suppose that there is just one $X$ field
and one $Y$ field.  Then the model we wish to study is defined
by the $N_i$-dependent terms in the superpotential
\beq
\mathcal{L}_N^W = \int d^2\theta \left( g\frac{X_{ij}}{M} L_i N_j H_u
+ g^\prime\frac{Y}{M}L_i N_i H_u + \ldots \right),
\label{newW}
\eeq
while the set of terms involving the rhd $N_i$ fields in the Kahler potential are
\beq
\mathcal{L}_N^K = \int d^4 \theta \left(
h  \frac{Y^\dagger}{M} N_i N_i +
{\tilde h} \frac{Y^\dagger Y}{M^2} N_i^\dagger N_i +
h_B \frac{Y^\dagger YX_{ij}^{\dagger} }{M^3} N_i N_j + \ldots \right).
\label{newK}
\eeq
The ellipses in Eqs.(\ref{newW}) and (\ref{newK}) stand either for
terms involving the replacement of $Y$ fields by $X_{ij}$ fields (with
obvious changes to $N_i$ flavour indices), or for terms higher
order in the $1/M$-expansion.  It is simple to check that
both types of additional term will lead to trivial or sub dominant
contributions not relevant for our discussion.
The Lagrangian displayed in (\ref{newW}) and (\ref{newK}) can be justified
with an R-Symmetry, where both hidden sector fields $X$ and $Y$ have R charge
$\frac{4}{3}$, $N$ has
R charge $\frac{2}{3}$, $E$ (rhd charged lepton superfield) has
R charge 2 and the remaining superfields have R charge equal to 0,
and the usual $R$-parity is assumed, with in addition  
$R_p(X)=R_p(Y)=+1$.  All dimensionless couplings $g,h$, etc, are
taken to be order one parameters.

Let us now suppose that after supersymmetry is broken in the hidden
sector at scale $m_I$ the field $Y$ acquires the
following $F$ and $A$-component vacuum expectation values,
\bea
&&\langle Y\rangle_F = F_Y = f_Y m_I^2 \nonumber\\
&&\langle Y\rangle_A = A_Y = 0 ,
\label{yvev}
\eea
while the field $X$ acquires the $F$ and $A$-component expectation
values:
\bea
&&\langle X_{ij}\rangle_F = F_{Xij} = 0 \nonumber\\
&&\langle X_{ij}\rangle_A = A_{Xij}= a_{Xij} m_I .
\label{xvev}
\eea
Here $f_{Y}$ and $a_{Xij}$ are order one parameters, and the zero
entries for $A_Y$ and $F_X$ can be replaced by
non-zero values $A_Y\ll m_I$, and $F_X \ll m_i^2$ without significant
change.  We will often re-write the scale $m_I$ in terms of the
gravitino mass $m_{3/2}=m_I^2/M$, and the reduced Planck mass $M$.

Before we proceed to analyze the consequences of the above effective
Lagrangian for the light neutrino masses and mixings, a
comment is in order concerning the assumption that $\vev{F_X}_{ij} \ll m_I^2$.
As is well-known, typical supergravity mediation of susy breaking
has difficulties with FCNC and CP-violation constraints unless there is
a high degree of degeneracy among the squark and slepton soft masses
of different generations (we here ignore the
possibility of alignment mechanisms which are typically
much harder to implement).  Such degeneracy is accommodated in our
model when $\vev{F_X}_{ij} \ll m_I^2$.  In this paper we will not be
concerned
about the detailed origin of this high-level of degeneracy, but take it
as a phenomenologically necessary assumption. As we will argue below,
in this case a simple prediction for the structure of the
light neutrino mass matrix can result.

\section{Neutrino and sneutrino masses at tree level}

Given Eqs.(\ref{newW}) and (\ref{newK}), the relevant terms in the
Lagrangian after the $X$ and $Y$ fields gain their expectation values
are thus
\beq
\mathcal{L}= \int d^2\theta \biggl( \lambda_{ij} L_i N_j H_u +
M_N N_i N_i \biggr) +
A \tilde{L}_i \tilde{n}_i h_u +
B^2_{ij} \tilde{n}_i \tilde{n}_j + \ldots,
\label{relevant}
\eeq
where $\tilde{n}_i$ are the rhd sneutrino fields, $\tilde{L}_i$
is the lhd slepton doublet, $h_u$ is the up-type
Higgs scalar doublet, and the omitted terms include the usual
soft scalar mass terms. First note that the
effective neutrino Yukawa coupling in the
superpotential of Eq.(\ref{relevant}) is suppressed in magnitude by
a factor of $(m_{3/2}/M)^{1/2} \sim 10^{-7} - 10^{-8}$,
\beq
\lambda _{ij} = g a_{Xij} \sqrt{\frac{m_{3/2}}{M}} ,
\label{yukawa}
\eeq
and gains its flavour structure from the $\vev{X_{ij}}_A$ expectation
value,
and in addition, the scale of the rhd neutrino masses is lowered to the
$\tev$ scale
\beq
M_N = h f_Y m_{3/2} .
\eeq
Second, there exists a
$\tev$-scale, but flavour diagonal, trilinear scalar $A$-term
\beq
A=g^\prime f_Y m_{3/2}.
\label{Aterm}
\eeq
Finally there is a small
but significant rhd sneutrino lepton-number violating $B$-term with
coefficient
\beq
B^2_{ij} = h_B f_Y^2 a^s_{Xij}\sqrt{\frac{m_{3/2}^{5}}{M}}
\label{Bterm}
\eeq
of magnitude $B^2_{ij} \sim ({\rm few}\times 100\mev)^2$ and flavour
structure related to that of the neutrino Yukawa coupling
($a^s_{Xij}$ denotes the symmetric part of $a_{Xij}$).

After electroweak symmetry breaking, with Higgs expectation values
$\vev{H_u^0} = v\sin\beta$ and $\vev{H_d^0}=
v\cos\beta$, the effective Lagrangian Eq.(\ref{relevant}) implies
that at tree level in the neutrino sector, after integrating out the
$\tev$-scale rhd neutrinos, there is generated
a light neutrino mass matrix of see-saw type:
\beq
(m_\nu^{\rm tree})_{ij} = - {v^2 \sin^2\beta\over M_N}
 \lambda_{ik}^T \lambda_{kj} \sim \frac{v^2}{m_{3/2}} \frac{m_{3/2}}{M} .
\label{tree}
\eeq
If one takes $M=2\times 10^{18}\gev$, $v=174\gev$, and all $a_{Xik}$,
$f_Y$, $g$ and $h$'s to be of order 1, this gives a contribution to
the light neutrino mass matrix of order $10^{-5} \ev$, certainly too
small to generate the required $\Delta m^2_{23}$ and $\Delta m^2_{12}$.

Note, however, that since
the operators that give rise to the Yukawa and rhd Neutrino mass terms
are higher-dimension non-renormalizable operators, a more appropriate
estimate of the couplings $g$ and $h$ might be the values
found by applying the so-called `naive-dimensional analysis' (NDA)
methodology which assumes that
the cutoff $M$ is bounded by the UV strong coupling scale of the
non-renormalizable theory, and in which geometrical factors of
$1/(4\pi)^2$ which enter loop calculations are taken into
account\cite{Manohar:1983md}.  (It is known that NDA works
well for estimating the coefficients of the higher-dimensional
operators in the low-energy chiral-Lagrangian description of QCD.)
A simple calculation shows that
the NDA estimates for the sizes of the couplings $h$ and $g$
which set the rhd neutrino mass and lhd-rhd-Higgs Yukawa coupling are
$h\simeq 1$ and $g \simeq 4\pi$
(and a UV strong coupling scale $\Lambda$ related to the reduced Planck mass
as $\Lambda \simeq 4\pi M$). This leads to an improved estimate
of the size of the tree (see-saw) contribution to the light neutrino
masses
\beq
(m_\nu^{\rm tree})_{ij} \simeq
- {v^2 \sin^2\beta\over M} {g^2\over h}
\simeq 10^{-3} \ev .
\label{bigtree}
\eeq
Alternatively, the large mass scale $M$ might be the string or
susy-GUT scale, thus similarly increasing the estimate for
$m_\nu^{\rm tree}$.  In any case, we will argue in the
next section that a small tree-level term of size Eq.(\ref{bigtree})
can sometimes be an interesting perturbation to the {\em dominant}
one-loop contribution to the light neutrino mass matrix.

Turning to the sneutrino sector of the theory the effective Lagrangian
Eq.(\ref{relevant}) implies, after electroweak symmetry
breaking, a 12~by~12 mass matrix that mixes the lhd and rhd sneutrinos
and their conjugates (here, for simplicity, assuming $A$ and $B^2$ real),
\beq
\left(\begin{array}{cccc}
  M_L^{2}\delta_{ij} & A v\sin\beta\delta_{ij} & 0 & 0 \\
   A v\sin\beta\delta_{ij} & M_R^2\delta_{ij} & 0 & B^2_{ij} \\
 0 & 0 & M_L^{2}\delta_{ij} & A v\sin\beta\delta_{ij} \\
 0 & B^2_{ij} & A v\sin\beta\delta_{ij} & M_R^2\delta_{ij}
\end{array}\right) ,
\label{snumasses}
\eeq
where we work in the
basis $({\tilde\nu}^*,{\tilde n},{\tilde\nu},{\tilde n}^*)$,
$M_L^2=m_L^2+m_Z^2\cos(2\beta)/2$ is the lhd
sneutrino mass arising from the usual soft mass $m_L^2$ and electroweak
breaking terms, and $M_R^2=m_R^2+M_N^2$ is the total rhd sneutrino
mass including the soft mass-squared. (Our notation is that Latin
indices run from 1 to 3, so e.g., the ${\tilde n}$ are indexed as
$i+3$, and sneutrino mass eigenstates are labeled by Greek indices
$\alpha,\beta=1,\ldots,12$. ) This mass matrix is diagonalised
by a unitary rotation of the form
\beq
U = \left(\begin{array}{cccc}
  V/\sqrt{2} & 0 & V/\sqrt{2} & 0 \\
   0 &  V/\sqrt{2} & 0 & V/\sqrt{2}\\
 -V/\sqrt{2} & 0 & V/\sqrt{2} & 0 \\
   0 &  - V/\sqrt{2} & 0 & V/\sqrt{2}
\end{array}\right)
\left(\begin{array}{cccc}
  \cos\phi & \sin\phi & 0 & 0 \\
   -\sin\phi &  \cos\phi & 0 & 0 \\
 0 & 0 & \cos\phi & \sin\phi \\
0 & 0 & -\sin\phi & \cos\phi
\end{array}\right)
\label{Umatrix}
\eeq
where $\tan 2\phi = 2 Av\sin\be/(M_L^2 - M_R^2)$, and $V$ is the
3~by~3 matrix that diagonalises $B_{ij}^2$.  The expression
Eq.(\ref{Umatrix}) is correct up to small terms of order
$B^2/(M_L^2 - M_R^2)$.  The corresponding
sneutrino mass eigenvalues fall into 2 groups
of almost 3-fold degenerate complex states.

\section{Structure of light neutrino masses at 1-loop}

\begin{figure}
  \begin{center}
  \epsfig{file= 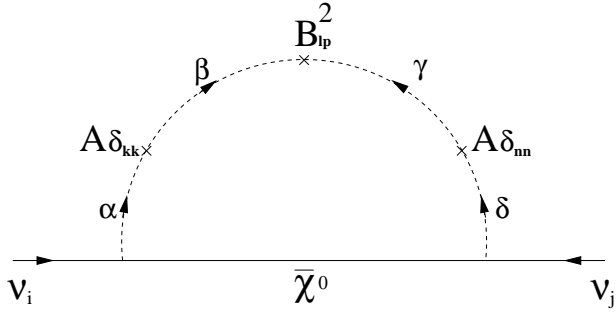}
  \caption{The dominant contribution to $m_{\nu}$.}
  \label{loop}
  \end{center}
\end{figure}

The rhd sneutrino interaction Eq.(\ref{Bterm}) gives rise to a radiative
contribution to the light neutrino masses, illustrated in Fig.~\ref{loop},
that, for $\vev{Y}_F \sim m_I^2$ and $\vev{X_{ij}}_A \sim m_I$,
{\em dominates} over the tree level contribution
arising from the the mixing of the TeV-mass rhd neutrinos with the lhd
neutrino states.
In detail, the contribution to the light Majorana neutrino mass
matrix from the diagram in Fig.~\ref{loop} is given by,
\beq
m_{\nu,ij}^{\rm loop} =
\frac{\chi_x A^2 B_{\ell p}^2 v^2 \sin^2\beta }{16\pi^2} U_{\alpha,
i+6} U^\dagger_{\alpha,k+6} U_{\beta,k+3} U^\dagger_{\beta,\ell+3}
U_{\gamma, p+3} U^\dagger_{\gamma,n+3} U_{\delta, n+6} U^\dagger_{\delta,j+6}
L(\alpha,\beta,\gamma,\delta,x) ,
\label{mnuinitial}
\eeq
where all repeated indices are summed over.
In this expression, $\chi_x$ is a factor that depends on the
exchanged neutralino, $U$ is the sneutrino mixing matrix,
$x=1,...,4$ denote the neutralino mass eigenstates, and
$L(\alpha,\beta,\gamma,\delta,x)$, is a totally symmetric
function of the sneutrino and neutralino masses that arises
from the momentum integral in Figure~\ref{loop}.

Because of the small effective Yukawa coupling,
Eq.(\ref{yukawa}), between the higgsinos, lhd neutrinos, and rhd
sneutrinos, in practice only the bino and zino components of
the neutralino are
exchanged across the bottom of the loop. Thus only these states
need to be expanded in terms of the 4 neutralino mass eigenstates,
leading to
\beq
\chi_{x}=\frac{2M_{Z}^{2}}{v^{2}}\biggl| N_{x1}^* \sin\theta_w -
N_{x2}^* \cos\theta_w \biggr|^2 M_{\chi_{x}}
\eeq
where $N_{xI}$ is the standard neutralino mixing matrix, and
$M_{\chi_{x}}$ are the neutralino masses.

A transparent and elegant form for the light neutrino mass matrix
results if we consider the situation in which the $A$-term mixing
is small.  This greatly simplifies the
flavor structure of the result Eq.(\ref{mnuinitial}). (The $V$
matrices in $U$ of Eq.(\ref{Umatrix}) cancel by virtue of
$VV^\dagger=1$.)  Moreover, in this limit the loop factor
$L$ also simplifies,
\bea
L & = & {1 \over M_R^6 (r-1)^3 (r-x)^2 (1-x)^2 }
\biggl[ (1-r)(1+r-2x)(1-x)(r-x) +  \biggl(2r^2-x^3(1+r)
\nonumber \\
& +& 2x^2(1+r^2 +r)\biggr)\log r - x\biggl( \log(r/x) + r\log(x^3 r) -
r^2\log(x^3/r^4) + r^3 \log x\biggr)\biggr] ,
\eea
where $r\equiv M_L^2/M_R^2$ and $x\equiv M_{\chi_x}^2 /M_R^2$.
Therefore in this limit of small sneutrino mixing the
1-loop contribution to the light neutrino mass matrix becomes
\beq
(m_\nu^{\rm loop})_{ij}=\sum_{x}\frac{\chi_{x}
A^2  B_{ij}^2 v^2 \sin^2\be }{(4\pi)^2}
L(M_L^2,M_R^2,M_R^2,M_L^2,M_{\chi_{x}}^2) .
\label{mnuloop}
\eeq

The most important feature of the result Eq.(\ref{mnuloop}) is
that the overall scale of the contribution $m_\nu^{\rm loop}$ is
naturally of the correct size to account for atmospheric neutrino
oscillations (as is also true of the models of
Refs.\cite{Arkani-Hamed:2000kj,Borzumati:2000ya}). To see this
explicitly it is useful to consider the simple case in which all
the lhd and rhd sneutrinos and the neutralinos are approximately
equal to a common mass scale, $m_{\rm susy}$, giving \beq L \simeq
-{1\over 12 m_{\rm susy}^6} . \label{loopfactor2} \eeq which leads
to a one-loop contribution of magnitude \beq m_\nu^{\rm loop} \sim
\mu \equiv {\al_w \over 96 \pi} {m_I^9 v^2\over M^5 m_{\rm
susy}^5} \simeq 10^{-2} \ev - 10^{-1} \ev \label{magnitude} \eeq
depending on the precise magnitude of the $A$ and $B$ terms. In
addition, in our model, the flavour structure of this dominant
one-loop contribution to the light neutrino mass matrix is
determined directly and entirely by the rhd sneutrino
lepton-number violating $B$-term, $B_{ij}^2$, which is in turn
generated by the $\vev{X_{ij}}_A$ expectation value. Moreover, as
claimed earlier, the one loop result dominates over the tree level
see-saw contribution. It is useful to define the (small) parameter
$\ep$ as the ratio of magnitudes of the tree-level see-saw
contribution Eq.(\ref{bigtree}) to the above 1-loop contribution.

It is also interesting to consider the regime in which the rhd
sneutrino states are heavy compared to the neutralino and lhd
sneutrino states, $r\simeq x \ll 1$. In this case the loop factor
is approximated by the expression \beq L \simeq \frac{(x-r)+
x\log(r/x)}{M_R^6 (r-x)^2}\simeq -\frac{1}{2M_R^4 M_L^{\prime 2}}
~. \label{loopfactor1} \eeq Taking $M_L\sim M_\chi\sim m_{\rm
susy}$ and scaling the $A$ and $B$ terms relative to their natural
values, $A_0^2\sim m_{\rm susy}^2$, and $B_0^2\sim m_{\rm susy}^2
(m_{\rm susy}/M)^{1/2}$ given by Eqs.(\ref{Aterm}) and
(\ref{Bterm}), leads to \beq m_\nu^{\rm loop}  \simeq {m_{\rm
susy}^{7/2} M_Z^2\over 16\pi^2 M_R^4 M^{1/2}} \left({A^2\over
A_0^2}\right) \left({B^2\over B_0^2}\right) . \eeq Assuming
$m_{\rm susy}\sim 300\gev$, and $M_R\sim 1\tev$ this gives
$m_\nu^{\rm loop} \sim 0.02\ev (A^2/A_0^2) (B^2/B_0^2)$ showing
that $M_R$ cannot be much heavier than $1\tev$ unless the scale of
the MSSM superpartners is uncomfortably high.

In either case the final structure of the light neutrino mass
matrix is in total
\beq
(m_\nu^{\rm tot})_{ij} = \mu (a_X^s + \ep a_X^T a_X)_{ij} .
\label{mnustructure}
\eeq
with the scale set by $\mu\sim 0.1\ev - 0.01\ev$, and $\ep$ in the range
$\ep\sim 10^{-2} - 10^{-4}$.
An attractive feature of this structure is that it allows us in a simple
way to account for the hierarchy between the solar and atmospheric neutrino
mass-squared splittings.  The atmospheric $\Delta m^2$ can arise from
the one-loop contribution, while the tree-level correction leads to the
small $\Delta m^2_{\rm solar}$ splitting, the hierarchy being
entirely due to the small dynamical parameter $\ep\sim 10^{-2}$.
Of course it is possible that the hierarchy instead arises entirely
from the flavour structure of the dominant
$\mu a_X^s$ term, the tree-level perturbation being insignificant, but
this leads to quite traditional neutrino flavour models, so we here focus
upon the new possibility.

As a simple example of a model along these lines consider the
situation in which the matrix $a_X$ corresponds to the
leading-order form for an inverted hierarchy model \beq a_X =
\pmatrix{ 0 & 1 & a \cr 1 & 0 & 0 \cr a & 0 & 0  \cr } . \eeq
Therefore in this case the full light neutrino mass matrix
including both loop and tree contributions has the form \beq
m_\nu^{\rm tot} = \mu \pmatrix{ (1+a^2)\ep & 1 & a \cr 1 & \ep &
a\ep \cr a & a\ep & a^2\ep  \cr }. \label{mnuexample} \eeq This
mass matrix has one zero eigenvalue, and two massive eigenvalues
$m_\pm \simeq  \mu(\sqrt{a^2 +1} \pm \ep)$.  Therefore to
accommodate the oscillation data requires $(1+a^2)\mu^2\simeq
2\times 10^{-3}\ev^2$, while the solar oscillation data requires
$4(1+a^2)^{1/2} \mu^2 \ep \simeq 7\times 10^{-5}\ev^2$.  For
$a\sim 1$ this gives \beq \mu^2 \simeq 10^{-3} \ev^2, \qquad \ep
\simeq 10^{-2} \eeq comfortably of the sizes expected in this
model. Moreover, since $m_\nu^{\rm tot}$ is diagonalised by first
performing a 23 rotation, $R_{23}(\th)$, with angle
$\th=\tan^{-1}(a)$, and then a 12 rotation, $R_{12}(\phi)$, with
$\phi=\pi/4$, we see that for $a\sim 1$ the atmospheric and solar
angles will both be close to maximal.  Specifically, recalling
that the physical neutrino mixing (MNS) matrix includes a unitary
matrix $V_L$ from the rotation of the charged leptons to a basis
where the charged lepton mass matrix is diagonal and real we find
a MNS matrix of the form

\beq V_{\rm MNS} = V_L R_{23}^{T}(\th) R_{12}^{T}(\phi) = V_L
\pmatrix{ c_\phi & s_\phi & 0 \cr -c_\th s_\phi & c_\th c_\phi &
s_\th \cr s_\th s_\phi & -s_\th c_\phi & c_\th  \cr } . \eeq Under
the not unreasonable assumption that the mixing angles in from the
charged lepton sector are small, $V_L$ only slightly perturbs the
above structure, thus leading to almost maximal physical
atmospheric and solar mixing angles, and a small $\th_{e3}$ angle.

Alternatively, light neutrino masses with a normal hierarchy can
be easily generated if $a_X$ contains an antisymmetric piece, $a_X^a$.
In particular consider the forms,
\beq
a_X^s=\pmatrix{
0 & 0 & 0 \cr
0 & b & b \cr
0 & b & b \cr }\qquad
a_X^a=\pmatrix{
0 & c & d \cr
-c & 0 & e \cr
-d & -e & 0 \cr }
\label{normalhierarchy}
\eeq
Substitution of these matrices into Eq.(\ref{mnustructure}) with
reasonable (and not fine-tuned) ${\cal O}(1)$ values for the
parameters $b$, $c$, $d$, and $e$, and values of the dynamically
determined parameters $\mu^2 \simeq {\rm few}\times 10^{-3} \ev^2$, and
$\ep \simeq 10^{-2}$, leads to a light neutrino mass matrix, which
when diagonalised produces mass squared differences within the
experimental bounds for a normal hierarchy.
Moreover, again under the assumption that the real mixing angles in
$V_L$ are small this leads to large physical atmospheric and solar
mixing angles, and a small $\th_{e3}$ angle.  More generally, if
the matrix $a_X$ is not real as in the above examples, but
contains large phases then it is simple to generate successful
models in which the one-loop term gives $\Delta m^2_{\rm atm}$
while the perturbing tree term leads to $\Delta m^2_{\rm solar}$.

\section{Comments and conclusions}

In this paper we have further analyzed a class of
models first introduced by Arkani-Hamed
et~al~\cite{Arkani-Hamed:2000bq,Arkani-Hamed:2000kj},
and Borzumati et al~\cite{Borzumati:2000mc,Borzumati:2000ya},
in which the light neutrino masses are a result of
higher-dimensional supersymmetry-breaking terms.  The
mechanism is closely related to the Giudice-Masiero
mechanism for the MSSM Higgs $\mu$ parameter, and in particular leads
to TeV-scale rhd neutrino and sneutrino states, that are in principle
accessible to direct experimental study, unlike traditional
see-saw mechanisms.  A second difference is that the dominant
contribution to the light neutrino mass matrix (which is of Majorana
type) is a one-loop term induced by a lepton-number violating and
supersymmetry breaking $B$-term for the sneutrino states
that is naturally present in the model.  In this letter we have
focused upon the simplification and analysis of the flavour structure
of this general class of models, and have found that simple
predictions for the light neutrino mass matrix are possible.  In
addition we have found that the subdominant tree-level `see-saw'
contribution may lead to interesting perturbations of the leading
one-loop-induced flavour structure, possibly generating the smaller
$\Delta m_{\rm solar}^2$.

In this paper we have not explored the important issues of the
possible collider and cosmological tests of our models.  In broad
structure the implications of our models are similar to those
already analyzed in
Refs.\cite{Arkani-Hamed:2000bq,Arkani-Hamed:2000kj,
Borzumati:2000mc,Borzumati:2000ya}.  In particular one expects the
$A$-term interactions in our model to lead to interesting
possibilities for production and decay of the TeV-scale rhd
sneutrino states, including the possibility of anomalous Higgs
decays. Another intriguing possibility is that the rhd sneutrino
states could be the dark matter \cite{Hall:1998ah}, or that, when
CP-violation in the neutrino sector is taken into account, dark
matter with dominant inelastic interactions with matter results
\cite{Smith:2001hy}. In a future publication, \cite{leptogen}, we
show that the class of models analyzed here naturally lead to a
very attractive and successful theory of TeV-scale resonant
leptogenesis, developing from the earlier work of
\cite{resonant}.

\vskip 0.05in
\begin{center}
{\bf Acknowledgments}
\end{center}
\vskip0.05in
We wish to thank Thomas Hambye, Graham Ross and
Lotfi Boubekeur for discussions. SW is
supported by PPARC Studentship Award PPA/S/S/2002/03530.


\begin{thebibliography}{99}

%\cite{Maltoni:2004ex}
\bibitem{Maltoni:2004ex}
M.~Maltoni,
%``Status of neutrino oscillations. I: The three-neutrino scenario,''
arXiv:hep-ph/0401042.
%%CITATION = HEP-PH 0401042;%%

%\cite{Apollonio:1999ae}
\bibitem{Apollonio:1999ae}
M.~Apollonio {\it et al.}  [CHOOZ Collaboration],
%``Limits on neutrino oscillations from the CHOOZ experiment,''
Phys.\ Lett.\ B {\bf 466} (1999) 415
[arXiv:hep-ex/9907037].
%%CITATION = HEP-EX 9907037;%%

%\cite{Ahmed:2003kj}
\bibitem{Ahmed:2003kj}
S.~N.~Ahmed {\it et al.}  [SNO Collaboration],
 %``Measurement of the total active B-8 solar neutrino flux at the Sudbury
%Neutrino Observatory with enhanced neutral current sensitivity,''
arXiv:nucl-ex/0309004.
%%CITATION = NUCL-EX 0309004;%%

%\cite{Eguchi:2002dm}
\bibitem{Eguchi:2002dm}
K.~Eguchi {\it et al.}  [KamLAND Collaboration],
 %``First results from KamLAND: Evidence for reactor anti-neutrino
%disappearance,''
Phys.\ Rev.\ Lett.\  {\bf 90} (2003) 021802
[arXiv:hep-ex/0212021].
%%CITATION = HEP-EX 0212021;%%

%\cite{Toshito:2001dk}
\bibitem{Toshito:2001dk}
T.~Toshito  [Super-Kamiokande Collaboration],
%``Super-Kamiokande atmospheric neutrino results,''
arXiv:hep-ex/0105023.
%%CITATION = HEP-EX 0105023;%%

\bibitem{seesaw} 
M. Gell-Mann, P. Ramond and R. Slansky, 
in {\it Supergravity}, edited by P. van Nieuwenhuizen and D. Freedman, 
(North-Holland, 1979), p.~315; 
S.L. Glashow, in Quarks and Leptons, Carg\`ese, eds. M. L\'evy et al., 
(Plenum, 1980, New-York), p. 707; 
T. Yanagida, in {\it Proceedings of the Workshop on the Unified Theory 
and the Baryon Number in the Universe}, edited by O. Sawada and
A. Sugamoto (KEK Report No.~79-18, Tsukuba, 1979), p.~95; 
R.N.~Mohapatra and G. Senjanovi\'{c}, Phys. Rev. Lett. {\bf 44}, (1980) 912. 


%\cite{Arkani-Hamed:2000bq}
\bibitem{Arkani-Hamed:2000bq}
N.~Arkani-Hamed, L.~J.~Hall, H.~Murayama, D.~R.~Smith and N.~Weiner,
%``Small neutrino masses from supersymmetry breaking,''
Phys.\ Rev.\ D {\bf 64}, 115011 (2001)
[arXiv:hep-ph/0006312].
%%CITATION = HEP-PH 0006312;%%

%\cite{Arkani-Hamed:2000kj}
\bibitem{Arkani-Hamed:2000kj}
N.~Arkani-Hamed, L.~J.~Hall, H.~Murayama, D.~R.~Smith and N.~Weiner,
%``Neutrino masses at v**(3/2),''
arXiv:hep-ph/0007001.
%%CITATION = HEP-PH 0007001;%%

%\cite{Borzumati:2000mc}
\bibitem{Borzumati:2000mc}
F.~Borzumati and Y.~Nomura,
%``Low-scale see-saw mechanisms for light neutrinos,''
Phys.\ Rev.\ D {\bf 64}, 053005 (2001)
[arXiv:hep-ph/0007018].
%%CITATION = HEP-PH 0007018;%%


%\cite{Borzumati:2000ya}
\bibitem{Borzumati:2000ya}
F.~Borzumati, K.~Hamaguchi, Y.~Nomura and T.~Yanagida,
%``Variations on supersymmetry breaking and neutrino spectra,''
arXiv:hep-ph/0012118.
%%CITATION = HEP-PH 0012118;%%

\bibitem{related}
Y.~Grossman and H.~E.~Haber,
%``Sneutrino mixing phenomena,''
Phys.\ Rev.\ Lett.\  {\bf 78} (1997) 3438 [arXiv:hep-ph/9702421]:
%%CITATION = HEP-PH 9702421;%%
F.~Borzumati, K.~Hamaguchi and T.~Yanagida,
%``Supersymmetric seesaw model for the (1+3)-scheme of neutrino masses,''
Phys.\ Lett.\ B {\bf 497} (2001) 259
[arXiv:hep-ph/0011141]:
%%CITATION = HEP-PH 0011141;%%
R.~Kitano,
%``Small Dirac neutrino masses in supersymmetric grand unified theories,''
Phys.\ Lett.\ B {\bf 539} (2002) 102 [arXiv:hep-ph/0204164]:
%%CITATION = HEP-PH 0204164;%%
J.~A.~Casas, J.~R.~Espinosa and I.~Navarro,
%``New supersymmetric source of neutrino masses and mixings,''
Phys.\ Rev.\ Lett.\  {\bf 89} (2002) 161801
[arXiv:hep-ph/0206276]:
%%CITATION = HEP-PH 0206276;%%
R.~Arnowitt, B.~Dutta and B.~Hu,
%``Yukawa textures, neutrino masses and Horava-Witten M-theory,''
arXiv:hep-th/0309033:
%%CITATION = HEP-TH 0309033;%%
S.~Abel, A.~Dedes and K.~Tamvakis,
%``Naturally small Dirac neutrino masses in supergravity,''
arXiv:hep-ph/0402287.
%%CITATION = HEP-PH 0402287;%%


%\cite{Nilles:1983ge}
\bibitem{Nilles:1983ge}
See e.g., H.~P.~Nilles,
%``Supersymmetry, Supergravity And Particle Physics,''
Phys.\ Rept.\  {\bf 110}, 1 (1984).
%%CITATION = PRPLC,110,1;%%


%\cite{Giudice:1988yz}
\bibitem{Giudice:1988yz}
G.~F.~Giudice and A.~Masiero,
%``A Natural Solution To The Mu Problem In Supergravity Theories,''
Phys.\ Lett.\ B {\bf 206}, 480 (1988).
%%CITATION = PHLTA,B206,480;%%


%\cite{Manohar:1983md}
\bibitem{Manohar:1983md}
A.~Manohar and H.~Georgi,
%``Chiral Quarks And The Nonrelativistic Quark Model,''
Nucl.\ Phys.\ B {\bf 234}, 189 (1984).
%%CITATION = NUPHA,B234,189;%%


%\cite{Hall:1998ah}:
\bibitem{Hall:1998ah}
L.~J.~Hall, T.~Moroi and H.~Murayama,
%``Sneutrino cold dark matter with lepton-number violation,''
Phys.\ Lett.\  {\bf B424}, 305 (1998)
[hep-ph/9712515].
%%CITATION = HEP-PH 9712515;%%


%\cite{Smith:2001hy}
\bibitem{Smith:2001hy}
D.~R.~Smith and N.~Weiner,
%``Inelastic dark matter,''
Phys.\ Rev.\ D {\bf 64}, 043502 (2001)
[arXiv:hep-ph/0101138].
%%CITATION = HEP-PH 0101138;%%


\bibitem{leptogen}
T.~Hambye, J.~March-Russell, and S.~West, arXiv:hep-ph/0403183


%\cite{Hambye:2001eu}
\bibitem{resonant}
A.~Pilaftsis,
%``CP violation and baryogenesis due to heavy Majorana neutrinos,''
Phys.\ Rev.\ D {\bf 56} (1997) 5431 [arXiv:hep-ph/9707235]:\\
%%CITATION = HEP-PH 9707235;%%
T.~Hambye,
%``Leptogenesis at the TeV scale,''
Nucl.\ Phys.\ B {\bf 633}, 171 (2002)
[arXiv:hep-ph/0111089].
%%CITATION = HEP-PH 0111089;%%



\end{thebibliography}
\end{document}